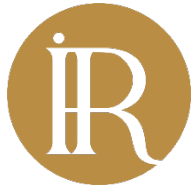

# Competing Visions of Ethical AI: A Case Study of OpenAI

*Melissa Wilfley\*, Mengting Ai\*, Madelyn Rose Sanfilippo*

## Abstract

**Introduction.** AI Ethics is framed distinctly across actors and stakeholder groups. We report results from a case study of OpenAI analysing ethical AI discourse.

**Method.** Research addressed: How has OpenAI's public discourse leveraged 'ethics', 'safety', 'alignment' and adjacent related concepts over time, and what does discourse signal about framing in practice? A structured corpus, differentiating between communication for a general audience and communication with an academic audience, was assembled from public documentation.

**Analysis.** Qualitative content analysis of ethical themes combined inductively derived and deductively applied codes. Quantitative analysis leveraged computational content analysis methods via NLP to model topics and quantify changes in rhetoric over time. Visualizations report aggregate results. For reproducible results, we have released our code at https://github.com/famous-blue-raincoat/AI_Ethics_Discourse.

**Results.** Results indicate that safety and risk discourse dominate OpenAI's public communication and documentation, without applying academic and advocacy ethics frameworks or vocabularies.

**Conclusions.** Implications for governance are presented, along with discussion of ethics-washing practices in industry.

\* Equal Contribution.

## Introduction

Artificial intelligence (AI) is advancing at a pace faster than the vocabularies and institutions meant to govern it (Cath, 2018; Lazar & Nelson, 2023). General-purpose systems such as large language models (LLMs), image generators and multimodal agents now circulate across work, politics and everyday life, producing sociotechnical dynamics that high-level ethical principles struggle to capture. Widely cited frameworks organized around beneficence, non-maleficence, justice, autonomy and transparency (Floridi & Cowls, 2019; Jobin et al., 2019) offer broad guidance, yet scholars argue these commitments frequently collapse when confronted with infrastructures, incentives and usage paradigms that outpace institutional capacity (Mittelstadt, 2019; Morley et al., 2021).

We approach 'AI ethics' not as a checklist of principles but as a discursive and generative field: one that sets expectations, legitimizes actors, shapes policy debates, and structures organizational practice and labour roles. Given that governance infrastructures and technical standards are still under construction, language does constitutive work (Hajer, 1995; Fairclough, 1992). The words a leading AI developer chooses, or avoids, help define what counts as responsible practice, whose expertise matters, and which risks are foregrounded or sidelined (Stamboliev & Christiaens, 2025).

If, for example, ethics is invoked only rhetorically, it can function more as reputational cover than as a binding constraint (Bietti, 2020; Green, 2021). Language choices reallocate epistemic authority in a field where the stakes are high, and the rules are still being written.

Prior work maps ethical principles (Jobin et al., 2019), critiques corporate rhetoric (Green, 2021), and assesses governance frameworks (Papagiannidis et al., 2025; Morley et al., 2021). Few studies, however, have traced over time how a leading AI developer publicly deploys, or omits, explicit ethics vocabulary relative to adjacent terms such as *safety* and *alignment*. We address the following research question: *How has OpenAI's public discourse leveraged ethics, safety, alignment and adjacent concepts over time, and what does discourse signal about ethical framing in practice?*

## Background

### Principle-based AI ethics

AI ethics scholarship often centres on principle-based frameworks that translate ethical theory into high-level commitments for sociotechnical AI systems. Widely cited approaches converge around beneficence and non-maleficence, autonomy, justice and explicability/transparency (Floridi & Cowls, 2019; Jobin et al., 2019). These principles emphasize rights, fairness, and accountability as values to guide AI design and deployment. At the same time, scholars increasingly caution that principles can remain abstract or symbolic, without mechanisms for enforcement (Mittelstadt, 2019). More recent work highlights the persistent gap between principles and practice, showing that as AI systems scale, organizations struggle to embed such normative commitments into day-to-day development, evaluation and release processes (Ryan et al., 2021; Stahl, 2022).

### Principles-to-Practice

AI developers have recently rolled out 'responsible AI' or 'AI safety' programs, that attempt to operationalize ethics through risk management processes, evaluation benchmarks, incident reporting, and model governance. This translation has clear benefits in offering technical fixes (e.g. bias audits and fairness benchmarks: Raji et al., 2020; Barocas et al., 2023; model cards: Mitchell et al., 2019; datasheets: Gebru et al., 2021), creating internal governance processes (e.g. incident logs, risk tiers and review boards: Raji et al., 2020; Morley et al., 2021; Metcalf et al., 2021), and establishing common standards that allow developers to coordinate around shared goals (IEEE, 2019; Ryan et al., 2021; Morley et al., 2021). However, there are trade-offs associated. Toolkits tend to be either too general to be practically useful or too rigid to adapt to specific sociotechnical contexts, leaving a persistent principle-to-practice gap (Morley et al., 2021; Metcalf et al., 2021; Herzog & Blank, 2024). Scholars also warn that corporate calls for 'responsibility' can function as governance-lite, where vague signals, like publishing an ethics statement, can be substituted for substantive accountability (Mittelstadt, 2019; Metcalf et al., 2021; Green, 2021; Stahl et al., 2022).

### Ethics-washing and Reputational Ethics

AI ethics-washing (and related concepts) conceptually labels cases where 'ethics' becomes a reputational asset rather than a binding constraint (Bietti, 2020; van Maanen, 2022; Seele & Schultz, 2022). Ethics then collapses into a communications layer of values pages and principles documents, while real decision-making is structured by product roadmaps, release cycles, partnerships and labour priorities. Companies tend to selectively use procedural vocabularies such as 'safety', 'trust', and 'responsibility', while sidelining explicitly political or justice-oriented language that can appear more disruptive within institutional settings (Stamboliev & Christiaens, 2025; Greene et al., 2019; Green, 2021; Heilinger, 2022). Ethnographic and interview-based studies further reveal how 'ethics owners' institutionalize responsibility as a matter of process management rather than moral constraint (Metcalf et al., 2019), how AI practitioners feel responsibility displaced by powerful institutional actors (Orr & Davis, 2020), and how developers gain ethical expertise even as ethics debates are displaced outside their communities (Griffin et al., 2025). Hao (2025) argues that unenforceable ethical statements normalize secrecy and consolidation while masking departures from earlier commitments to openness and collaboration.

### Discourse

From sociotechnical and discourse perspectives, organizational language does constitutive work: it legitimizes actors, defines problems and solutions, and sets boundaries on what 'responsible development' can look like (Hajer, 1995; Fairclough, 1992). Technical processes and documents are not neutral, they script norms and values into industry practices (e.g., Shilton, 2018). Scholars have shown how corporate vocabularies shape ethical debates by institutionalizing responsibility as managerial process instead of a moral constraint (Metcalf et al., 2019), framing value statements to enable some conversations while foreclosing on justice-oriented alternatives (Greene et al., 2019) and narrowing ethical discourse toward techno-solutionism (Heilinger, 2022).

For example, Hao (2025) illustrates this discursive power at OpenAI. Altman framed companies as 'religions' (p. 9), Sutskever redefined 'openness' as an openness in spirit, and Brockman cast AI as a neutral actor serving humanity (p. 42). Executives further positioned 'safety' existentially, foregrounding catastrophic risk scenarios (*Our Updated Preparedness Framework*, 2025) while deflecting attention from nearer-term AI ethical concerns like labour displacement (U.S. Senate Committee on the Judiciary, Subcommittee on Privacy, Technology, and the Law., 2023; Hao, 2025). Such discourse shapes norms, privileging and delegitimizing certain risks and responsibilities.

## Method

We employed a mixed-methods approach to examine how OpenAI constructs and deploys ethics discourse in its public communications between December 2015 and July 2025. Our analysis combined computational text analysis to capture large-scale quantitative patterns with a qualitative audit to interpret the contextual nuances of ethics related terminology. This design traces both the frequency and the faming of 'ethics', 'safety' and adjacent concepts across OpenAI's web articles and publications.

### Case Selection

Four ex-ante criteria were identified for a suitable case for longitudinal analysis of AI ethics discourse: **1) Frontier AI development:** an organization developing and deploying general-purpose/foundational models at the technological frontier; **2) Ethics discourse participation:** the company has made public commitments and authored or co-authored recurring use of terms such as 'ethics', 'safety', 'alignment', and 'responsible AI' for at least five years; **3) Temporal traceability:**

a dated, publicly accessible archive of news, research posts and linked publications that enables time-series analysis; **4) Scale and Impact:** models surpassing the general-purpose AI threshold guidelines of $10^{23}$ FLOPs (approx. 1B parameters) as defined in the *EU Artificial Intelligence Act* (Regulation (EU) 2024/1689).

OpenAI uniquely satisfies all four criteria. Selecting a single, information-rich case allows us to combine large-scale computational measures with close qualitative auditing while holding organizational context constant.

### Corpus Construction and Preprocessing

To ensure that the analysis reflected discourse generated by OpenAI itself, we restricted our dataset to OpenAI-authored and co-authored materials published through its primary public communications channel (OpenAI.com) and its main research dissemination venue (arXiv), where OpenAI routinely releases preprints of its formal publications. Our corpus was collected from two sources to enable both quantitative and qualitative analysis:

**OpenAI Website Articles (n = 424):** We systematically collected all publicly available web articles from OpenAI's News (OpenAI research, 2025) and Research (OpenAI news, 2025) sections.

**Publications (n = 30):** To capture OpenAI's research register, we included OpenAI-authored and co-authored publications that the organisation either hosted directly or linked through its website. Because most publications surfaced on OpenAI.com also appeared on arXiv, we supplemented the site-hosted corpus by systematically identifying additional OpenAI-authored or co-authored publications. This set contains:

> **a) arXiv preprints (n = 25):** From 137 initially identified, 26 addressed ethics; one was excluded due to inaccessibility via OpenAI.com, yielding 25.
> **b) Site-hosted items (n = 5):** The website surfaced 43 publications via primary call-to-action (CTA) links—29 PDFs hosted on OpenAI's CDN and 14 HTML reports. Of these, only 3 CDN PDFs and 2 HTML reports contained a variation of the term *ethic* (e.g., ethic, ethics, ethical) and were not duplicates,

The total corpus includes **454 documents: 424 web articles and 30 publications** (25 arXiv preprints and 5 site-hosted items). While we analyse publications as a single category, we note their distinct provenance and, where relevant, report findings separately for arXiv and site-hosted outputs.

All texts were processed through a standardized pipeline: lowercasing, removal of punctuation, digits, and HTML artifacts, and lemmatization. Stopwords were removed using a standard English list augmented with corpus-specific terms (e.g., *openai*, *gpt*, *window*). This ensured conceptual consistency across corpora and minimized noise in downstream analysis. For web articles, paratextual material was excluded, such as footnotes, references, contributors and acknowledgements. For publications, only the title, subtitle, abstract and body text were included. This ensured the analysis focused on substantive discursive content instead of bibliographic or paratextual methods of ethics. The dataset reflects curated, English-language materials OpenAI elected to publish, representing a snapshot versus a complete archival record of all prior discourse.

### Quantitative Analysis

Our computational framework consisted of two complementary methods: (1) a targeted keyword-concept frequency analysis to track pre-defined concepts, and (2) an unsupervised topic modeling approach to discover emergent thematic structures.

The first pillar of our quantitative analysis aimed to quantify the prevalence and chronological shifts of known, pre-defined ethics and safety concepts.

**1) Concept Library Construction**: We developed a comprehensive keyword library organized by core concepts. We manually grouped related terms and their variations (e.g., grouping *safe*, *safety*, and *safely* under a unified SAFETY concept) to ensure that the analysis tracked high-level ideas rather than just individual words. The final library consisted of 75 core concepts.

**2) Frequency Tracking & Visualization**: After standardized preprocessing, we calculated concepts' annual frequencies. The results were visualized using heatmaps, which allowed for a clear comparison of concept prominence over time and revealed distinct patterns between different document types, such as official PDF reports and more informal web articles. This method provides a robust, quantitative baseline of how OpenAI discusses specific, pre-identified topics.

The second pillar moved beyond pre-defined keywords to discover the latent thematic structures and conceptual groupings. This bottom-up approach allows for the discovery of nuanced or unexpected themes.

1) **N-gram Vector Representation**: We extracted significant bigrams and trigrams from the preprocessed corpus to serve as proxies for core concepts. Each of these n-grams was then encoded into a high-dimensional vector using the all-MiniLM-L6-v2 SentenceTransformer model (Reimers & Gurevych, 2019), which captures their semantic meaning in context.

2) **Semantic Clustering:** We applied the HDBSCAN (McInnes et al., 2017) density-based clustering algorithm to these n-gram vectors. Unlike methods that require a pre-specified number of clusters (e.g. K-Means: Lloyd, 1982), HDBSCAN automatically determines the number of emergent clusters based on the density of the semantic space, providing contextual grouping of concepts.

3) **Manual Thematic Labeling & Visualization:** The resulting clusters, each containing a set of semantically related n-grams, were manually reviewed and assigned thematic labels (e.g., 'Reinforcement Learning Policy,' 'Public Data & Security Controls'). We generated visualizations to map thematic clusters and illustrate relationships, providing a discourse concept map over time.

## Qualitative Analysis

To further validate and contextualize computational findings, we conducted a systematic manual audit of OpenAI web articles that contained 'ethic-', were tagged 'Ethics & Safety' in Research and 'Safety' in News, and their primary call-to-action (CTA) linked publications. We also reviewed all OpenAI authored or co-authored arXiv preprints that contained the term 'ethic-'.

Each item was logged in a structured matrix that captured both metadata (document type, year, authorship, URL, tags, and source location on OpenAI.com, arXiv, or CDN) and ethics-related usage (e.g., whether 'ethic' appeared in the title, abstract, executive summary, section headers, diagrams, or body text). For posts with CTA links, the linked publication was downloaded and recorded in the matrix.

Web articles and publications were coded for explicit use of the term *ethics* as well as adjacent moral language (*moral*, *values*, *norms*). Coding tracked both frequency and framing, noting term location in the web article or publication.

The manually identified subset was cross-checked against the computational corpus to confirm coverage and resolve any discrepancies. This ensured that all items meeting the inclusion criteria were represented in both quantitative and qualitative analyses.

Additionally, we conducted exploratory check to ensure adequate coverage. First, we performed a landscape audit of other OpenAI.com properties using a targeted search query ('site:openai.com ethics') to capture results across the site and related subdomains. Second, we used the Internet Archive's Wayback Machine to spot-check historical snapshots of OpenAI's News and Research pages, focusing on changes to topic categories and tagging practices over time. ArXiv preprint titles were cross-referenced on OpenAI.com to identify official links. These supplementary steps were not exhaustive but provided additional confidence in the completeness of our dataset and helped identify edge cases where tagging practices had shifted, or where ethics discourse was applied.

Ambiguous or missing cases were reviewed collaboratively by at least two researchers before inclusion or exclusion. This collaborative process aimed to minimize bias and ensure consistency.

## Results

### Use of *Ethics* in OpenAI.com Web Articles

Across the 424 web articles in our OpenAI.com corpus, explicit references to *ethics* were rare. Only 16 articles (3.8%) contained a variation of the term (e*thic(s)*, *ethical*, *unethical*, *ethicist(s)*), making it one of the least frequently used vocabularies in the AI Ethics discursive dataset. By contrast, other ethics-adjacent terms appear far more frequently but undergo sharp temporal expansions. *Safe/safety* (lemmatized) peaks at nearly 700 mentions, over two orders of magnitude higher than *ethics*. *Risk* (386 in 2024) and *responsible* (~50 mentions annually after 2023) also surge, while *alignment* and *governance* emerge only after 2019, reflecting OpenAI's shift towards technical and institutional framings.

| Keyword | 2015 | 2016 | 2017 | 2018 | 2019 | 2020 | 2021 | 2022 | 2023 | 2024 | 2025 |
|---|---|---|---|---|---|---|---|---|---|---|---|
| alignment | 0 | 0 | 0 | 2 | 5 | 4 | 4 | 57 | 37 | 40 | 18 |
| benefit_humanity | 1 | 0 | 0 | 1 | 4 | 1 | 3 | 1 | 18 | 18 | 10 |
| ethics | 0 | 0 | 0 | 0 | 2 | 1 | 2 | 0 | 1 | 7 | 3 |
| governance | 0 | 1 | 0 | 0 | 1 | 0 | 0 | 0 | 33 | 18 | 5 |
| policy | 0 | 11 | 143 | 83 | 27 | 10 | 11 | 20 | 92 | 122 | 57 |
| responsible | 0 | 3 | 0 | 3 | 8 | 2 | 5 | 5 | 50 | 28 | 37 |
| risk | 0 | 2 | 0 | 11 | 19 | 9 | 12 | 45 | 203 | 386 | 199 |
| safe | 0 | 13 | 16 | 30 | 121 | 27 | 24 | 64 | 384 | 687 | 322 |
| trustworthy | 0 | 0 | 2 | 1 | 5 | 1 | 1 | 3 | 32 | 51 | 24 |

**Figure 1.** Annual counts for focal keywords on OpenAI.com web articles (tabular view). The table highlights the 2023-2024 peak for *safety* and *risk*, alongside the persistently low frequency of *ethics*.

Temporal patterns (Figure 1 & 2) are revealing. Mentions of *ethics* per year peak at just seven in 2024. By contrast, *safe/safety* reaches 687 mentions, and *risk* is mentioned 386 times in 2024. Other terms show delayed but notable growth—*alignment* is negligible until 2019 and then rises to 57 mentions by 2022, clustering around n-gram phrases such as *alignment problems*, *alignment research*, and *deliberative alignment*. *Governance* remains scarce until 2023-2024, when it registers a visible uptick. The aspirational phrase 'benefitting humanity' is modest (≤ 18), while *responsible* follows a similar trajectory with a low frequency until 2023, then spiking to ~50 mentions and remaining elevated through 2025.

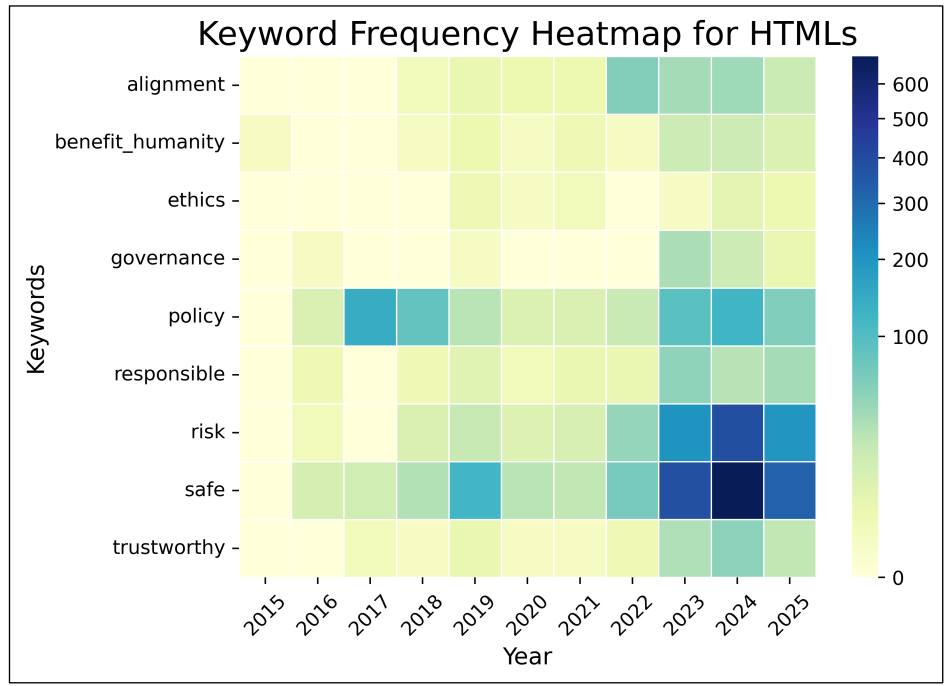

**Figure 2.** AI ethics keyword frequency heatmap of OpenAI.com web articles (2015–2025). The heatmap visualizes temporal trends, showing how *ethics* remains rare in discourse compared to growing prominence of concepts such as *safety*, *risk*, *alignment*, *responsibility*, *trustworthy* and *governance*.

Our qualitative audit shows that even within the 16 web articles that reference *ethics*, usage is minimal and largely superficial. 11 web articles mention the term only once, and the remaining 5 no more than three times. The highest concentration occurs in the 2024 announcement "OpenAI appoints Scott Schools as Chief Compliance Officer", where the *ethics* is used solely to describe Schools' prior role as 'Chief Ethics and Compliance Officer at Uber Technologies' (*OpenAI Appoints Scott Schools as Chief Compliance Officer*, 2024). Nearly all other references to ethics are in passing, often in stock phrases such as 'ethical standards' (*Pioneering an AI Clinical Copilot with Penda Health*, 2025) or 'ethical innovation' (*OpenAI's Commitment to Child Safety*, 2024), rather than a sustained analytic frame.

Qualitative analysis of OpenAI's web articles confirm this uneven distribution in tagging practices. As of July 2025, 52 Research articles were tagged 'Ethics & Safety' (*OpenAI Research*, 2025), yet only 7 of these used *ethics* in the body text. In the News section, 38 articles were tagged 'Safety' (*OpenAI News*, 2025), but only 3 used included *ethics*. Safety-tags are not evenly distributed across years: they appear in 2016, resurface in 2020, and then reoccur more consistently from 2022-2025. The number rises sharply in 2023 (13 articles, compared to just 2 in 2022), followed by 10 in 2024 and 11 by mid-2025. By contrast, 'Ethics & Safety' tags appear steadily from 2016-2025, but with little substantive use of *ethics* in the body.

Importantly, tagging conventions themselves have shifted over time. For instance, the article "Planning for AGI and beyond" (2023) was initially tagged 'Safety & Alignment' (2023; Wayback Machine) but now appears only under 'Safety'. This suggests that tagging functions less as a stable classification system and more as a dynamic metadata practice used as a form of discursive signalling, reflecting changes in how OpenAI curates and publicly frames AI development.

These trends point to a broader discursive narrowing in OpenAI's public communications. The website overwhelmingly foregrounds *safety* and *risk*, with annual frequencies in the hundreds compared to single-digit mentions of *ethics*. Consistent with our qualitative audit, when *ethics* does appear, it is typically peripherally, used as a single reference in a header or brief line in the article body such as 'uphold the highest ethical standards' (*Pioneering an AI Clinical Copilot with Penda Health*, 2025). During this time frame, *ethics* functions more rhetorically than substantively, its marginal presence standing in stark contrast to the hundreds of mentions of *safety* and *risk* that dominate OpenAI's public voice.

### OpenAI-Authored and Co-Authored Publications

Explicit references to *ethics* were far more prevalent across OpenAI-authored and co-authored publications, though still far from frequent. Out of 180 publications reviewed, 31 contained a variation of the term *ethic* (17.2%), more than four times higher than the 3.8% rate observed in OpenAI's web articles. The bulk of these came from arXiv (26/137, 19.0%), with a smaller share from site hosted materials (5/43, 11.6%).

Of the 31 texts that used *ethics* language, all but one were accessible via OpenAI's website. The final publications corpus therefore consists of 30 items. Within this set, references were typically minimal: 80% (24/30) mention the term three times or fewer, with only a handful exceeding five mentions. The most notable frequencies occur in "The Malicious Use of Artificial Intelligence" (2018, 9 mentions), "AI Safety Needs Social Scientists" (2019, 11 mentions), and "Toward Trustworthy AI Development" (2020, 12 mentions). Our qualitative audit shows that when *ethics* does appear, it usually falls into three recurring contexts: (1) external references, appendices, footnotes or datasets; (2) policy and guideline copy; or (3) appeals to societal norms and values. Like OpenAI's web articles, this is often located in headers, call-out boxes, diagrams or peripheral framing devices rather than sustained analytic engagement (see Table 1). Related terms such as *morals*, *values* and *norms* surfaced only sporadically, such as in stock phrases ('align with human values'; Mu et al., 2024), or references to social dynamics ('diffusion of bad norms'; OpenAI et al., 2024). Adjacent terms like *bias*, *harm* or *discrimination* rarely co-occur with *ethics and* were framed in technical or reputational registers rather than moral registers.

| Use context | Example |
| --- | --- |
| External references, appendices, footnotes, or datasets | Citing an AI ethics research dataset named 'ETHICS' (Burns et al., 2023) or referencing ethical guidelines in a collab workshop paper (Shoker & Reddie, 2023). |
| Policy/guideline content | Technical documents (like system cards or model specs) mention 'ethical' or 'unethical' behavior in example scenarios of disallowed content (*Sora system card*, 2024), to label misuse (e.g. forging docs as 'illegal or unethical' in a model reasoning example) (Guan et al., 2025). |
| Invoking societal norms/values | References to 'ethical standards' (*OpenAI's Approach to AI and National Security*, 2024) or rhetorical headers such as 'Continued safety, policy, and ethical alignment' (*Operator System Card*, 2025). |

**Table 1.** Where ethics language appears within publications.

Authorship patterns provide further context. Of 30 publications, 13 were co-authored (2018-2023, 2025) by OpenAI. Two of the three most-ethics dense texts were co-authored: "Toward Trustworthy AI Development" (2020; 12 mentions) and "The Malicious Use of AI" (2018; 9 mentions). This suggests that collaborations with academic and external partners create stronger incentives to explicate ethics.

Audience also matters. Most publications are available via arXiv, situating them in an academic register where at least some reference to *ethics* is expected. By contrast, OpenAI's website articles, targeted to broader audiences, largely displace *ethics* discourse in favour of *safety*. These patterns indicate that ethics is a discursive register activated most visibly in academic and collaborative contexts.

### Corpus-wide structure of OpenAI

Considering the entire corpus (424 web articles and 180 machine-readable publications), there are sharp contrasts in how OpenAI structures discourse temporally and structurally across public-facing and academic registers.

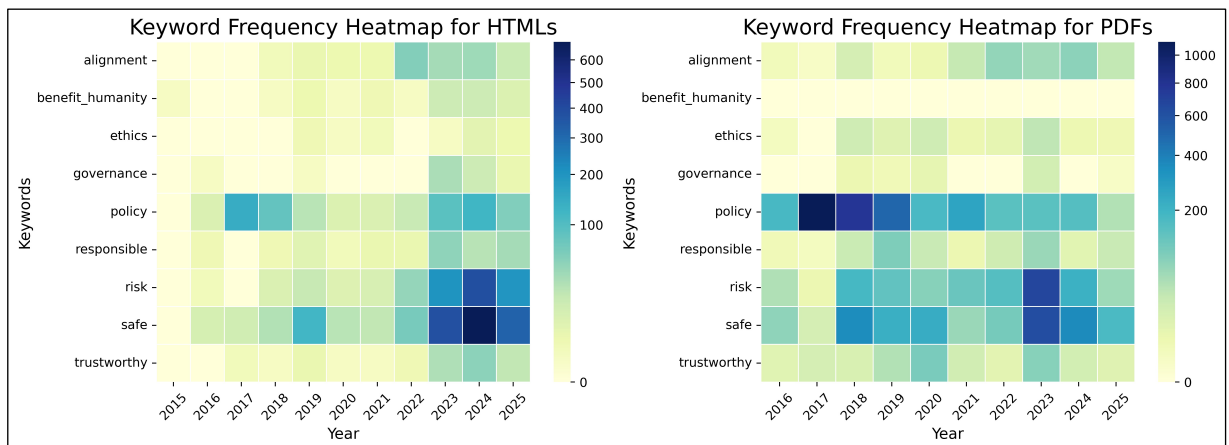

**Figure 3**. Keyword frequency heatmaps across OpenAI corpora (2015-2025). Left: web articles. Right: publications (PDFs and HTML). *Safety* and *risk* dominate both corpora, while *ethics* remains marginal throughout.

Temporal patterns, shown in Figure 3, highlight nine focal keyword frequencies across web articles (left) and publications (right). In web articles, *safety* and *risk* dominate, rising sharply after 2022 and peaking in 2024, while *ethics* remains extremely rare across the full decade. Terms such as *responsible* and *alignment* also surge after 2021, tracking OpenAI's pivot toward technical and governance-adjacent framings.

In publications, *policy* registers an early peak (2017-2019) before declining in relative prominence. *Safety* and *risk* remain consistently present, with visible spikes in 2019 and 2023. Publications reveal shifting foci: *policy* anchors discourse in the late 2010s, *safety* and *risk* become prominent by the early 2020s, and *alignment* emerges more recently as a technical focal point.

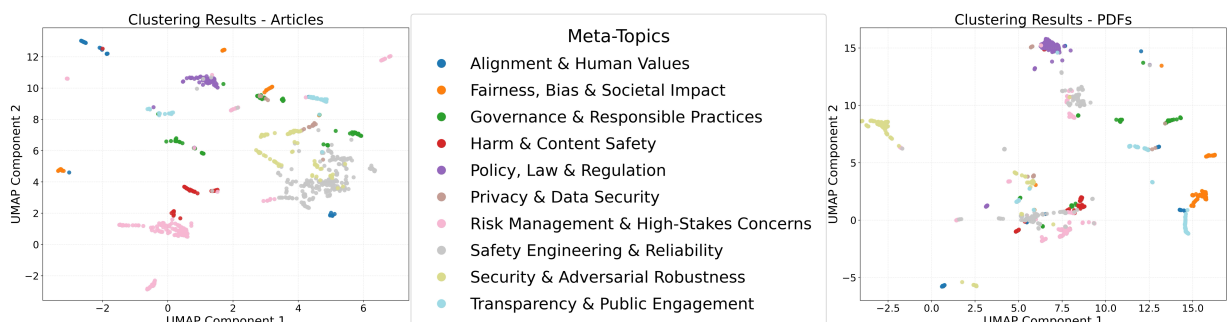

**Figure 4**. PCA clustering of OpenAI corpora. Left: web articles. Right: publications (PDFs and HTMLs.)

PCA clustering reveals a structural divergence between OpenAI's web articles and its publications (see Figure 4). The web articles illustrate integrated discussion where many topics overlap, creating a kind of 'melting pot'. In comparison, OpenAI-authored and co-authored publications show greater thematic specialization, with some topics being very distinct while others are still intertwined.

In the web articles, topics such as 'Safety Engineering & Reliability' (grey), 'Governance & Responsible Practices' (green), 'Security & Adversarial Robustness' (yellow), and 'Transparency & Public Engagement' (light blue) converge. This suggests that public communications fold diverse concerns into a single, accessible frame of 'AI Safety', with safety as the gravitational hub.

Publication clusters are more distinct: 'Policy, Law & Regulation' (purple) and 'Governance & Responsible Practices' (green) form clearly bounded regions, while 'Safety Engineering & Reliability' (grey), 'Risk Management & High-Stakes Concerns' (pink), and 'Harm & Content Safety' (red) overlap partially. Here, *safety* does not dominate but operates more as a proxy that interacts with *risk*, *harm*, *governance* across different subfields.

### Discursive Pivots: Safety and Policy

In the corpus-wide PCA, ethics never operates as an organizing category. OpenAI's discourse adheres around two pivots: safety, which consolidates and institutionalizes over time, and policy, which undergoes a semantic drift from technical reinforcement learning (RL) to governance vocabulary.

The trajectory of *safety* is clearest in the web corpus, where n-grams make visible a shift from scattered rhetorical fragments to systematic institutionalization. In the early years (2016–2018), mentions were rare ('concrete safety problem,' 'people safety'), compared with RL discourse. By 2019, however, *safety* had formalized, with OpenAI's Safety Gym platform benchmarks ('safety gym environment') anchoring technical usage while phrases like 'safety policy' and 'safety requirement' gestured toward governance. From 2020 to 2022, the vocabulary diversified into evaluation and compliance ('safety bounty,' 'safety incident'), before spiking five-fold in 2023–2024, when *safety* attached itself to *risk*, *governance*, and organizational frames ('safety risk,' 'safety security committee'). By mid-2025, the term remained saturated with specification language ('safety check,' 'safety testing'), emphasizing its role as a procedural proxy.

These granular shifts in OpenAI's web articles triangulate observations that *safety* scales flexibly across different registers. On the site, it acts as the gravitational hub that absorbs adjacent concerns into a unified 'AI safety' narrative. In publications, it connects specialized subfields, like risk management, harm/content safety, reliability, without anchoring them. Taken together, *safety* emerges not as a marker of *ethics* than but as a signal for compliance, governance, and evaluation.

*Policy* offers different insight. Both web articles and publications register an anomalous surge in 2017–2018 (143 and 200+ mentions, respectively), when *policy* was dominated by reinforcement learning usage ('policy gradient,' 'train policy,' 'policy head'). Mentions decline sharply in 2019–2022, then resurge in 2023–2024 (92 and 122 mentions on the site; 200+ in publications). During this later period, semantics shift as governance and compliance uses take hold, through phrases such as 'safety policy,' 'usage policy,' and 'policy proposals'. Heatmaps for both web articles and publications confirm that this is a corpus-wide pattern, *policy*'s trajectory reflects a broader reorientation.

## Discussion

### OpenAI's Public Discourse as a Signal about Ethics in Practice

Addressing our research question, as OpenAI absorbed *ethics* into a *safety* and *alignment* framing, it unified development by encouraging buy-in from technological teams, investors, and partners. However, it jeopardized societal and justice-oriented dimensions, diverging considerably from its initial mission. This transition aligned with increasing privatization and leveraging of IP to assert power and control over AI development. These changes, rooted in discourse, significantly impacted perceptions of what and who are critical and what or who is peripheral or expendable. For example, emphasis on 'safety' privileges research and staff that focus on technical aspects or risk and compliance, rather than a diverse array of values or ethical principles. Focus on 'alignment' centers ML and technosolutionism, alienating affected communities, civil society, social science experts.

### Implications

The case study of OpenAI highlights divergence between meaningful ethics dialogues and industry assurances on safety, trustworthiness, and human impact of AI. OpenAI increasingly omits an ethics vocabulary, which can quietly reshape AI governance practices on compliance, risk, and responsibility. Without ethically-grounded governance, AI oversight is sparse, addressing only technical safety or market-driven priorities.

OpenAI is a useful case study to understand how ethics-washing shapes norms and broader discourse, in addition to how it obscures and distracts from practice. Internally, there is also a significant impact of dialogue and framing on the design process, policy drafting, product safety processes, and priorities in team membership, reflecting a lack of coherent institutional ethics. Rhetoric actively redistributes epistemic authority in the AI ecosystem and governance. This highlights the need for transparency and accountability measures that are not subject to organizational dynamics or priorities; as evidenced by the reinstatement of Sam Altman as OpenAI's CEO on November 22, 2023 (Metz et al., 2023; *Sam Altman returns as CEO, OpenAI has a new initial board*, 2023), governance arrangements that are subject to pressure or capture are not institutionally sound.

### Future Directions

Given the significant shifts at OpenAI during this period of rapid AI innovation and adoption in ethics dialogue, parallel research ought to evaluate dialogue within the broader industry, considering trends by firm, market establishment, and across geopolitical divides. Further, researchers should empirically consider how these statements correspond with technical practices and compliance with evolving AI governance and regulation.

### Conclusions

This research highlights the shifts over time toward emphasis on safety, risk, and compliance, rather than nuanced ethos or meaningful social benefits. Results identified significant ethics washing, in line with recent research (e.g., Hao, 2025).

In evaluating OpenAI's public discourse and the associated signals about ethics in practice, there are broader implications around how language choices steer external accountability and impacts regulatory exposure and assurance posture. As such, a core takeaway from this case study is that governance and accountability for AI must be exogenously imposed and reflect nuanced ethical frameworks. Multistakeholder dialogue regarding AI ethics must be robust and sustained. Ephemeral and episodic events or partnerships are insufficient to ensure that human values and ethical concerns are centred in the development and deployment of AI innovations. While academics may coalesce around particular frameworks, such as FATE or FAccT, their nuanced considerations are not effectively transferred to industry practices or communications. Future research must compare this case to the broader industry and its impacts on society.

### Acknowledgements

The authors are grateful for the support of the School of Information Sciences at the University of Illinois at Urbana-Champaign and constructive feedback from reviewers and iConference 2026 program committee members.

## About the author(s)

**Melissa Wilfley** is a Ph.D. student in Information Sciences at the University of Illinois Urbana-Champaign (UIUC). She brings over 20 years professional experience in the technology sector and holds an A.B. in Sociocultural Anthropology from the University of California, Davis. Her research examines AI ethics, governance and tech labour dynamics in global contexts. She can be contacted at [wilfley2@illinois.edu](mailto:wilfley2@illinois.edu)

**Mengting Ai** is a Ph.D. student in Information Sciences at the University of Illinois Urbana-Champaign (UIUC). They received their M.S. in Computer Science from UIUC. Their research interests include AI ethics and large language models. They can be contacted at [mai10@illinois.edu](mailto:mai10@illinois.edu)

**Madelyn Rose Sanfilippo** is an Assistant Professor in the School of Information Sciences at the University of Illinois Urbana-Champaign. She received her Ph.D. from Indiana University and her research addresses governance in sociotechnical systems. She can be contacted at [madelyns@illinois.edu](mailto:madelyns@illinois.edu)